\documentclass[aps,prl,reprint,twocolumn,superscriptaddress,floatfix,nofootinbib,longbibliography]{revtex4-2}
\usepackage{graphicx,amsmath,amsfonts,amssymb,amsthm,xr}
\usepackage{epsfig}
\usepackage{tikz}
\usepackage{graphicx}
\usepackage{physics}
\usepackage{dcolumn}
\usepackage{bm}
\usepackage{ mathrsfs }
\usepackage{orcidlink}
\usepackage{algorithm}
\usepackage{algpseudocode}
\usepackage{float}

\usepackage{amsmath,amssymb}

\makeatletter
\renewcommand*\env@matrix[1][*\c@MaxMatrixCols c]{%
  \hskip -\arraycolsep
  \let\@ifnextchar\new@ifnextchar
  \array{#1}}
\makeatother

\usepackage{appendix} 

\begin{document}

\title{Krylov Complexity Meets Confinement}

\author{Xuhao Jiang}
\affiliation{Department of Physics and Arnold Sommerfeld Center for Theoretical Physics (ASC), Ludwig Maximilian University of Munich, 80333 Munich, Germany}
\affiliation{Technical University of Munich, TUM School of Natural Sciences, Physics Department, 85748 Garching, Germany}

\author{Jad C.~Halimeh${}^{\orcidlink{0000-0002-0659-7990}}$}
\affiliation{Department of Physics and Arnold Sommerfeld Center for Theoretical Physics (ASC), Ludwig Maximilian University of Munich, 80333 Munich, Germany}
\affiliation{Max Planck Institute of Quantum Optics, 85748 Garching, Germany}
\affiliation{Munich Center for Quantum Science and Technology (MCQST), 80799 Munich, Germany}
\affiliation{Department of Physics, College of Science, Kyung Hee University, Seoul 02447, Republic of Korea}

\author{N.~S.~Srivatsa${}^{\orcidlink{0000-0001-6433-450X}}$}
\email{srivatsa.nagara@mpq.mpg.de}
\affiliation{Max Planck Institute of Quantum Optics, 85748 Garching, Germany}
\affiliation{Munich Center for Quantum Science and Technology (MCQST), 80799 Munich, Germany}

\begin{abstract}

In high-energy physics, confinement denotes the tendency of fundamental particles to remain bound together, preventing their observation as free, isolated entities. Interestingly, analogous confinement behavior emerges in certain condensed matter systems, for instance, in the Ising model with both transverse and longitudinal fields, where domain walls become confined into meson-like bound states as a result of a longitudinal field-induced linear potential. In this work, we employ the Ising model to demonstrate that Krylov state complexity---a measure quantifying the spread of quantum information under the repeated action of the Hamiltonian on a quantum state---serves as a sensitive and quantitative probe of confinement. We show that confinement manifests as a pronounced suppression of Krylov complexity growth following quenches within the ferromagnetic phase in the presence of a longitudinal field, reflecting slow correlation dynamics. In contrast, while quenches within the paramagnetic phase exhibit enhanced complexity with increasing longitudinal field, reflecting the absence of confinement, those crossing the critical point to the ferromagnetic phase reveal a distinct regime characterized by orders-of-magnitude larger complexity and display trends of weak confinement.
Notably, in the confining regime, the complexity oscillates at frequencies corresponding to the meson masses, with its power-spectrum peaks closely matching the semiclassical predictions.

\end{abstract}

\maketitle

\textbf{\emph{Introduction.---}}Confinement, a hallmark of non-Abelian gauge theories such as quantum chromodynamics (QCD), refers to the phenomenon whereby fundamental excitations—quarks and gluons—cannot be isolated and instead form bound states such as mesons and baryons~\cite{Berges_review,QCD_review,Weinberg_book}. While originally formulated in the context of high-energy physics, confinement has also been shown to emerge in certain low-dimensional condensed-matter systems, where elementary excitations experience an effective linear potential that binds them into composite quasiparticles~\cite{Calabrese1,Calabrese2,Calabrese3,McCoy,Fangli2019,Lerose2019,Mussardo1,Mussardo2,Mussardo3,Mussardo4}. A number of experiments have explored various aspects of confinement, both in quasi-one-dimensional compounds~\cite{Wang2016,Morris2014,Grenier2015} and in quench experiments with cold-atom systems ~\cite{Langen_2016,Schneider2012,Greiner2002}.

A paradigmatic model exhibiting confinement is the one-dimensional Ising chain subjected to both transverse and longitudinal fields. The transverse-field Ising model is integrable and supports freely propagating domain-wall excitations; however, the addition of a longitudinal field breaks integrability and induces a linear confining potential that binds domain walls into meson-like bound states. As demonstrated by Kormos \textit{et al.}~\cite{Kormos2017}, this confinement profoundly alters the real-time dynamics, leading to suppressed correlation spreading and reduced entanglement growth, while generating oscillations characteristic of the meson mass spectrum. Notably, such confinement-induced dynamics have also been observed experimentally on an \texttt{IBM} quantum computer~\cite{Vovrosh2021}.

\begin{figure}
  \includegraphics[width=\linewidth]{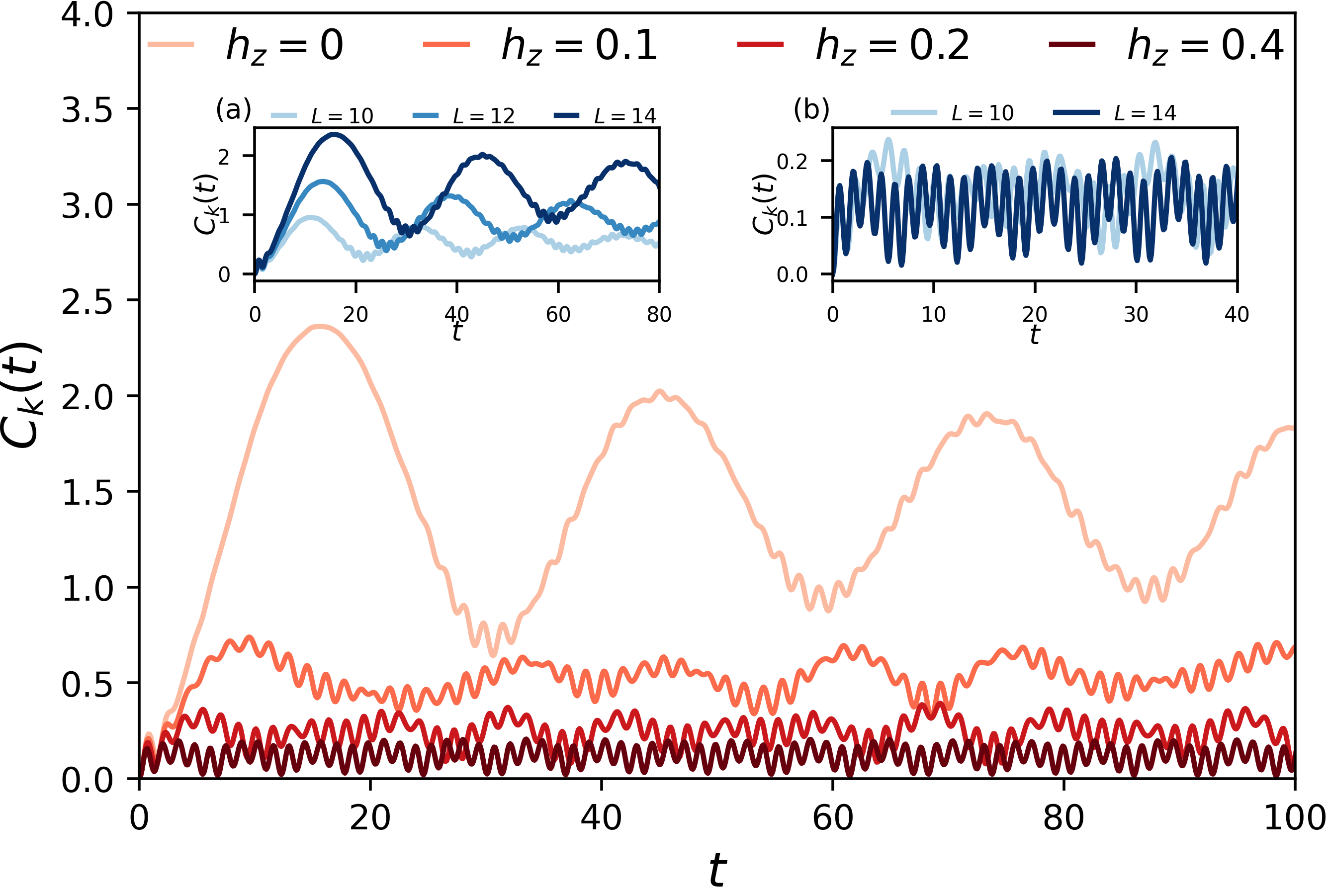}
  \caption{Krylov complexity $C_k(t)$ following a quench from a fully ferromagnetic initial state to $h_x = 0.25$ within the ferromagnetic phase, shown for various values of the longitudinal field $h_z$. A pronounced suppression in the complexity is observed upon introducing the longitudinal field. Inset (a) highlights marked finite-size effects in $C_k(t)$ at $h_z = 0$, while inset (b) illustrates the suppression of these effects for $h_z = 0.4$ as a consequence of confinement. For the main plot, we have considered a system of size $L = 14$.}
  \label{f1}
\end{figure}

These observations suggest that confinement places stringent constraints on the propagation of information and correlations in quantum many-body systems. Motivated by this, in this Letter we explore whether signatures of confinement can be identified through measures of quantum complexity. In particular, we study \textit{Krylov complexity}, which quantifies the spread of a time-evolved quantum state or an operator within the Krylov subspace generated by successive applications of the Hamiltonian or a Liovillian on an initial state or an operator~\cite{rabinovici}. The dynamics of complexity can be mapped to the motion of a quantum particle on a semi-infinite chain, allowing a geometric interpretation of the spread of the wavefunction~\cite{Vijay,Caputa2022}. Apart from its role in quantifying the spread of quantum information, Krylov complexity, defined for operators or states, has been shown to capture signatures of quantum chaos and thermalization in non-integrable closed and open systems~\cite{Parker2019,rabinovici,Hashimoto2023,Baggioli2025,Baggioli:2025knt,Chen2025,Alishahiha2025,Alishahiha2025b,gautam2023,NANDY20251,Cao_2021,Camargo2024,ChenLia2025,Rabinovici2022,Caputa:2025ucl,Rabinovici:2020ryf,Bhatta2022,Bhatta2025,Bhatta2023,Tanay2025,Sri2024,medina202,medina2025}. While Krylov operator complexity has been proposed as an order parameter for confinement–deconfinement transitions in large-$N$ gauge theories~\cite{Anegawa2024}, how the signatures of real-time confinement manifest within the Krylov space of quantum states following a quench remains largely unexplored.

We study the dynamics of Krylov state complexity in the quantum Ising model and show that it serves as a sensitive probe of confinement, as illustrated in Fig.~\ref{f1}. Following a quench within the ferromagnetic phase, we find that the growth of complexity is strongly suppressed upon introducing even a weak longitudinal field, reflecting the inhibited spread of quantum correlations due to the emergence of a confining potential. In the absence of this field, Krylov complexity exhibits pronounced oscillations with larger amplitude, consistent with the ballistic propagation of correlations inside the light cone. In stark contrast, quenches within the paramagnetic phase exhibit the opposite trend: the Krylov state complexity increases with the longitudinal field, reflecting dynamics akin to quantum chaotic behavior indicating absence of confinement. Quenches across the critical point, from the paramagnetic to the ferromagnetic phase, display yet another distinct behavior, where the complexity is several orders of magnitude larger than in the other two cases and shows trends of weak confinement. Furthermore, the oscillatory structure of dynamics of Krylov complexity for quenches within the confining phase encodes detailed spectral information of the confined regime. Remarkably, the peaks in its power spectrum correspond precisely to the meson bound-state masses, in excellent agreement with semiclassical predictions.

\textbf{\emph{Ising model and confinement.---}}
We consider the prototypical model for confinement, an Ising model with transverse ($h_x$) and longitudinal ($h_z$) fields given by the following Hamiltonian.

\begin{align}\label{Ising}
\hat{H}=-J\sum_{j=1}^{N}\big[\hat{\sigma}^z_{j}\hat{\sigma}^z_{j+1}+h_x\hat{\sigma}^{x}_{j}+h_z\hat{\sigma}_j^{z}\big].
\end{align}
In this work, we set $J=1$ throughout and use periodic boundary conditions. In the absence of a longitudinal field ($h_z = 0$), the model maps onto a system of non-interacting fermions and exhibits a quantum phase transition at the critical transverse field strength $h_x = 1$. For $h_x < 1$, the system is ferromagnetically ordered, characterized by two degenerate ground states related by a global $\mathbb{Z}_2$ symmetry. The low-energy excitations in this regime correspond to freely propagating domain walls (kinks) that separate regions of opposite magnetization and move ballistically in time. Kormos \textit{et al.}~\cite{Kormos2017} showed that a small longitudinal field $h_z$ breaks the $\mathbb{Z}_2$ symmetry, lifting the ferromagnetic degeneracy and inducing a linear confining potential that binds domain walls into meson-like excitations. The resulting bound states form a discrete spectrum whose confinement strength grows with $h_z$. Following a quench from a ferromagnetic state, it was demonstrated that the magnetization $\langle \hat{\sigma}^z(t) \rangle$ exhibits oscillations even for weak fields, while the entanglement entropy rapidly saturates, reflecting confinement dynamics. Interestingly, the oscillations in magnetization and entanglement encode information about the underlying meson spectrum, which manifests as distinct peaks in their power spectra~\cite{Kormos2017,Knaute2023}. 

These measures are useful for detecting confinement, however they primarily capture the spread of spatial correlations rather than how the state itself evolves. In particular, entanglement entropy is sensitive to the choice of bi-partition, highlighting the need for complementary probes such as complexity, which quantifies the global spreading of the quantum state in Hilbert space.

\begin{figure}
    \includegraphics[width=\linewidth]{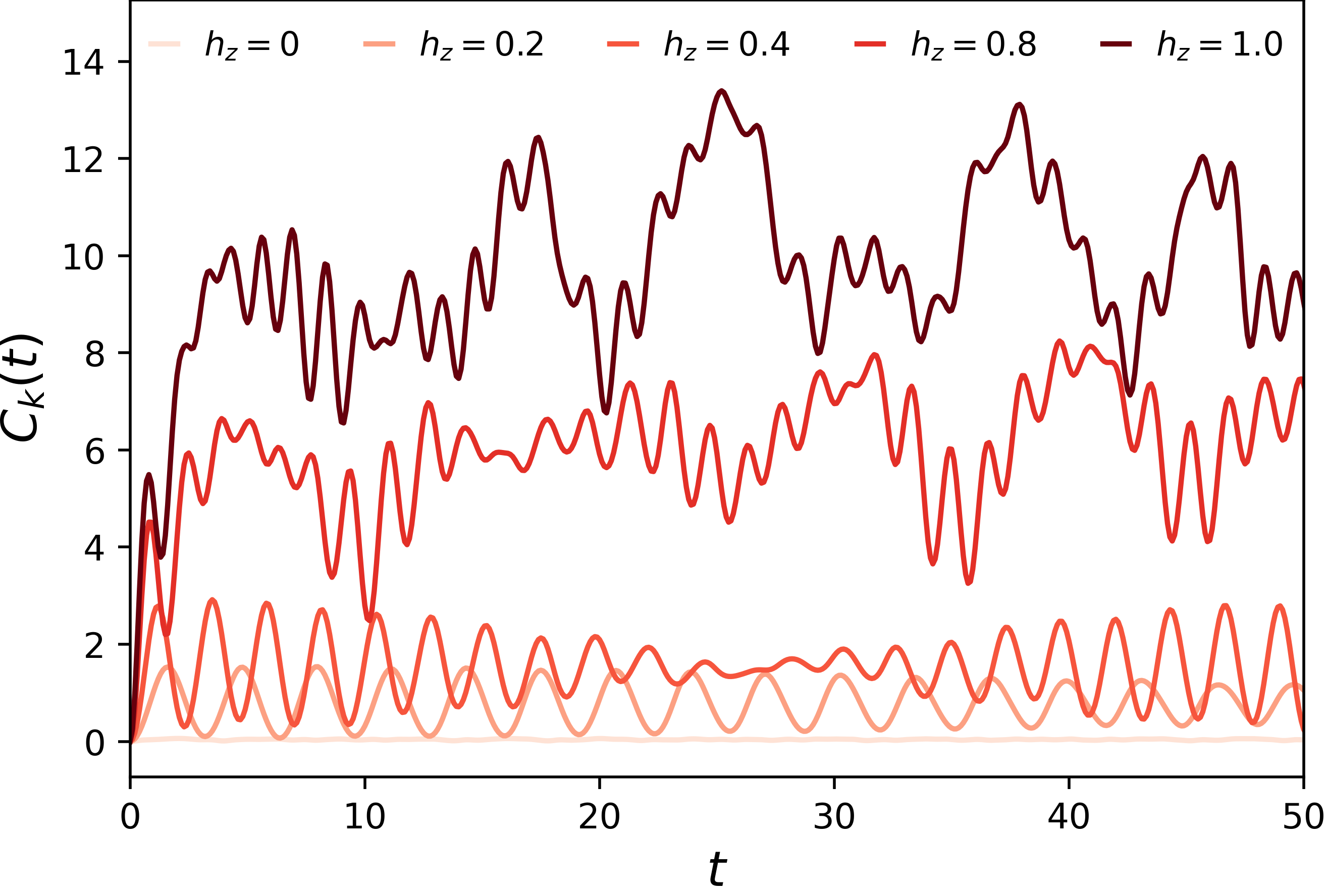}
    \caption{Krylov complexity $C_k(t)$ following a quench from a paramagnetic initial state at $h_x=2$ to $h_x = 1.75$ within the paramagnetic phase, shown for different values of the longitudinal field $h_z$. The observed increase in the amplitude of complexity with increasing $h_z$ indicates the absence of confinement, sharply differing from the quench response within the ferromagnetic phase. The system size used is $L = 14$.}
    \label{f2}
\end{figure}

\textbf{\emph{Krylov complexity.---}}We explore the imprint of confinement on the dynamics of quantum complexity, specifically through the lens of Krylov state complexity~\cite{rabinovici}. To this end, we construct the Krylov basis by successive applications of the Hamiltonian \(\hat{H}\) on an initial state \(\ket{\Psi_0}\), forming the subspace $\mathcal{K}_n(\hat{H}, \ket{\Psi_0}) = \mathrm{span}\{\ket{\Psi_0},\, \hat{H}\ket{\Psi_0},\, \dots,\, \hat{H}^{n-1}\ket{\Psi_0}\}$.
An orthonormal set \(\{\ket{K_j}\}\) is obtained via the \textit{Lanczos algorithm}~\cite{lanczos1950,parlett1998}, which tridiagonalizes the Hamiltonian as $\hat{H}\ket{K_n} = \alpha_n\ket{K_n} + \beta_{n+1}\ket{K_{n+1}} + \beta_n\ket{K_{n-1}}$,
where \(\alpha_n\) and \(\beta_n\) are the diagonal and off-diagonal Lanczos coefficients. Expanding the time-evolved state $\ket{\Psi(t)} = \sum_n \psi_n(t)\,\ket{K_n}$ and substituting into the Schrödinger equation gives $i\,\partial_t \psi_n(t) = \alpha_n \psi_n(t) + \beta_n \psi_{n-1}(t) + \beta_{n+1} \psi_{n+1}(t)$, which is equivalent to a tight-binding model on a semi-infinite chain with amplitudes \(\psi_n(t)\). The Krylov complexity is defined as
\begin{equation}
    C_k(t) = \sum_n n\,|\psi_n(t)|^2,
\end{equation}
quantifying the spread of the wavefunction in Krylov space. We provide more details of the Lanczos algorithm employed in this work in the SM~\cite{SM1}.

\textbf{\emph{Quench Regimes and Complexity Growth.---}}We begin by studying the dynamics of Krylov complexity following a quantum quench within the ferromagnetic phase ($h_x < 1$), starting from a fully polarized initial state, and compare cases with and without the confinement field $h_z$. For $h_z = 0$, the model is integrable, and as shown in Fig.~\ref{f1}, the Krylov complexity exhibits oscillations whose amplitude increases with system size (see inset), consistent with the results of Kormos \textit{et al.}~\cite{Kormos2017}, where longitudinal spin–spin correlations display a clear light-cone spreading. Introducing a finite longitudinal field $h_z$ breaks integrability and leads to a pronounced suppression of the complexity, with the suppression becoming stronger as $h_z$ increases, revealing clear signatures of confinement. The oscillation frequency increases while its amplitude decreases, reflecting the arrested dynamics of mesonic excitations. Physically, this occurs because the mesons acquire large effective masses, and the quench energy is insufficient to set them in motion, allowing only their creation at rest. This picture is further supported by the system-size independence of the Krylov dynamics even at long times (see inset), consistent with the absence of correlation spreading.

\begin{figure}
    \includegraphics[width=\linewidth]{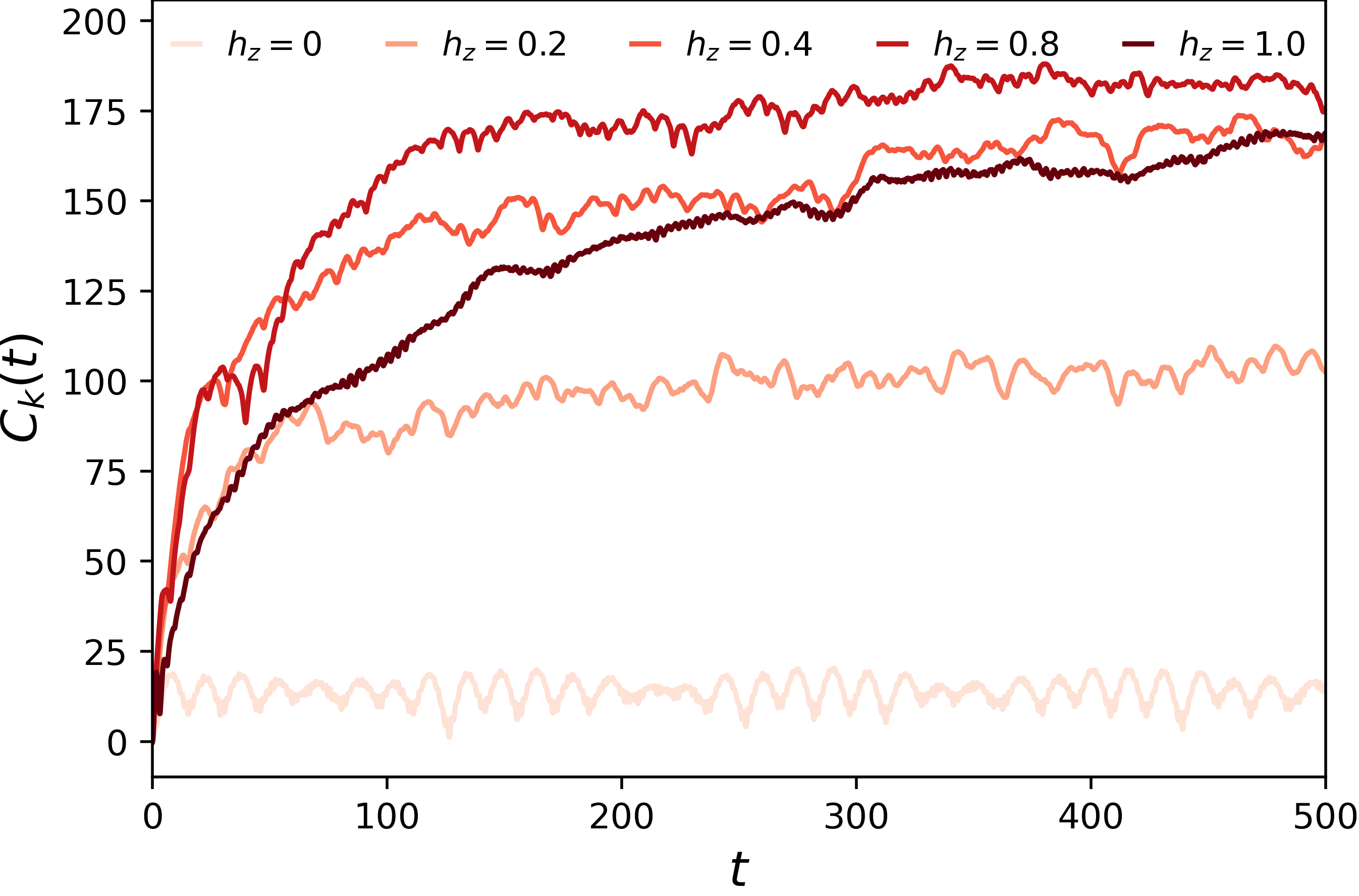}
    \caption{Krylov complexity $C_k(t)$ following a quench from a paramagnetic initial state at $h_x=2$ to $h_x = 0.25$, crossing the quantum critical point ($h_x = 1$), shown for various values of the longitudinal field $h_z$. The complexity initially increases with $h_z$ before eventually decreasing at larger field strengths which might indicate weak confinement. Notably, its magnitude is several orders higher than that observed for quenches performed entirely within the paramagnetic or ferromagnetic phases. The system size used is $L = 14$.
}
    \label{f3}
\end{figure}

We now turn to quenches within the paramagnetic phase ($h_x > 1$) starting from a paramagnetic state at $h_x=2$, as shown in Fig.~\ref{f2}. In sharp contrast to the ferromagnetic regime, the dynamics without a longitudinal field $h_z$ displays a markedly suppressed Krylov complexity with only small-amplitude fluctuations, characteristic of free-fermion behavior and the regular oscillations observed at small $h_z$ arise from the relatively small quench applied. Increasing $h_z$ leads to a rapid growth in the overall amplitude of the complexity, reflecting enhanced quantum chaotic dynamics arising from interactions. The paramagnetic phase thus remains non-confining and does not host bound meson-like excitations, underscoring the qualitative difference in its dynamical behavior compared to the ferromagnetic case

Next, we investigate the growth of complexity for large quenches across the quantum critical point at $h_x = 1$. Starting from a paramagnetic initial state at $h_x=2$, we quench into the ferromagnetic phase to $h_x = 0.25$. As shown in Fig.~\ref{f3}, the resulting complexity grows to values orders of magnitude larger than those observed for quenches performed within either the paramagnetic or ferromagnetic phases. The large amplitude of the complexity in this case arises because the quench across the critical point excites a broad continuum of modes and generates strong delocalization of the state in Krylov space, reflecting highly non-perturbative dynamics. Unlike intra-phase quenches, the post-quench dynamics in this case exhibits random fluctuations superimposed on the overall growth once the longitudinal field is introduced. Similar to the behavior observed in paramagnetic quenches, the complexity initially increases with the strength of the longitudinal field but eventually shows a distinct turnover, beyond which it begins to decrease. This crossover may signal the emergence of weak confinement, although the signature remains subdued, likely due to the finite propagation velocities of the mesons induced by the large quench amplitude.

\begin{figure*}
    \includegraphics[width=\linewidth]{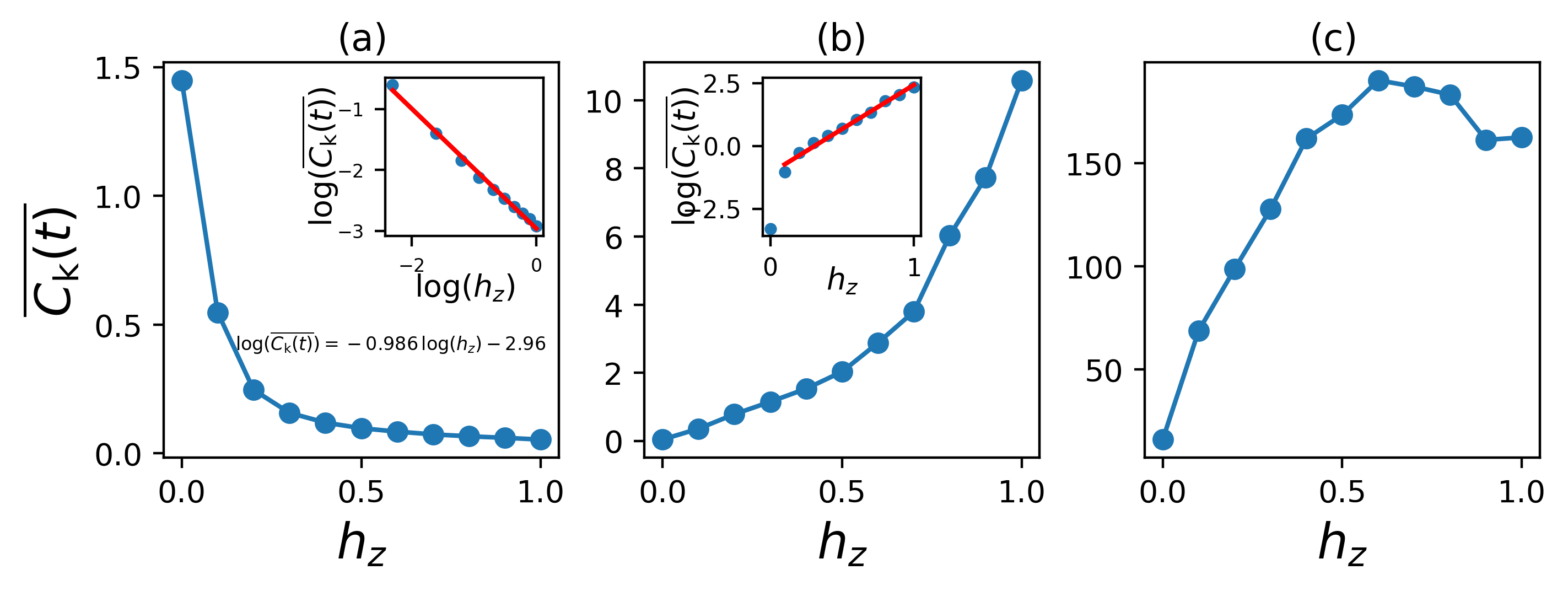}
    \caption{Dependence of the long-time–averaged Krylov state complexity $\overline{C_k(t)}$ on the confining longitudinal field $h_z$ for the three quench protocols discussed in the main text. (a) Quench entirely within the ferromagnetic phase, from a fully ferromagnetic initial state to $h_x = 0.25$. The inset shows $\overline{C_k(t)}$ plotted on a log--log scale, revealing an approximately linear dependence with slope close to $-1$, indicative of the behavior, $\overline{C_k(t)} \propto 1/h_z$. (b) Quench entirely within the paramagnetic phase, from a paramagnetic initial state at $h_x = 2$ to $h_x = 1.75$. The inset displays $\overline{C_k(t)}$ on a log--linear scale, where the asymptotically linear behavior suggests an exponential enhancement of the Krylov complexity with increasing longitudinal field $h_z$. (c) Quench from a paramagnetic initial state at $h_x = 2$ to $h_x = 0.25$, crossing the quantum critical point at $h_x = 1$. For all the cases, the system size used is L=14.
}
    \label{f4}
\end{figure*}

Lastly, to understand how the Krylov complexity scales with the strength of the confining longitudinal field $h_z$, we compute the long-time–averaged Krylov complexity, defined as
\begin{equation}
\overline{C_k(t)} = \lim_{T \to \infty} \frac{1}{T} \int_{0}^{T} C_k(t)\, dt \, .
\end{equation}
This quantity characterizes the asymptotic growth of the many-body wavefunction in Krylov space by averaging out short-time fluctuations and transient dynamics. We analyze $\overline{C_k(t)}$ for all three quench protocols discussed above. The results are shown in Fig.~\ref{f4}. For quenches conducted entirely within the ferromagnetic phase, the data indicate an inverse scaling of the complexity with the longitudinal field, $\overline{C_k(t)} \propto h_z^{-1}$. In contrast, for quenches within the paramagnetic phase, the behavior is qualitatively different. In this case, the time-averaged Krylov complexity exhibits an exponential enhancement with increasing $h_z$, signaling a rapid growth of state complexity driven by the longitudinal field. Finally, for quenches across the quantum critical point from the paramagnetic phase into the ferromagnetic phase where the dependence of $\overline{C_k(t)}$ on $h_z$ does not reveal a clear scaling form. Nevertheless, we observe that the Krylov complexity initially increases with $h_z$, followed by a bending and eventual saturation to a plateau at larger field strengths. This saturation may be attributed to weak confinement effects, which ultimately limit further growth of the state in Krylov space at long times. 

\textbf{\emph{Meson masses and Krylov spectroscopy.---}}As discussed, introducing a small longitudinal field in the ferromagnetic phase confines the otherwise free domain-wall excitations into meson-like bound states. The meson masses can be estimated semiclassically by modeling two fermions in a linear confining potential~\cite{Rutkevich2008,Kormos2017}, described by the effective two-body Hamiltonian $H_\text{two-body} = \epsilon(p_1) + \epsilon(p_2) + \chi |x_2 - x_1|$, where $p_{1,2}$ are the canonical momenta conjugate to the coordinates $x_{1,2}$. The single-particle dispersion after mapping Eq.~\eqref{Ising} to a system of spinless fermions (via a Jordan--Wigner transformation) is $\epsilon(p) = 2J\sqrt{1 + h_x^2 - 2h_x\cos p}$, with a confining potential $\chi = 2Jh_z\bar{\sigma}$, where $\bar{\sigma} = (1 - h_x^2)^{1/8}$ is the spontaneous magnetization of the ferromagnetic ground state. Applying Bohr--Sommerfeld quantization to this relative motion yields the bound state energy levels and consequently one can extract the meson masses. For instance, at $h_x = 0.25$ and $h_z = 0.2$, this semiclassical analysis predicts two mesons with masses $m_1 = 4.025J$ and $m_2 = 4.702J$. Decreasing the longitudinal field gives rise to more meson bound states (more details in the SM~\cite{SM1}).

When the system is quenched within the ferromagnetic phase, the Krylov state complexity $C_k(t)$ displays pronounced oscillations that directly arise from the presence of confined mesons. These oscillations therefore encode the energy scales and characteristic signatures of the bound modes.

\begin{figure}
    \includegraphics[width=\linewidth]{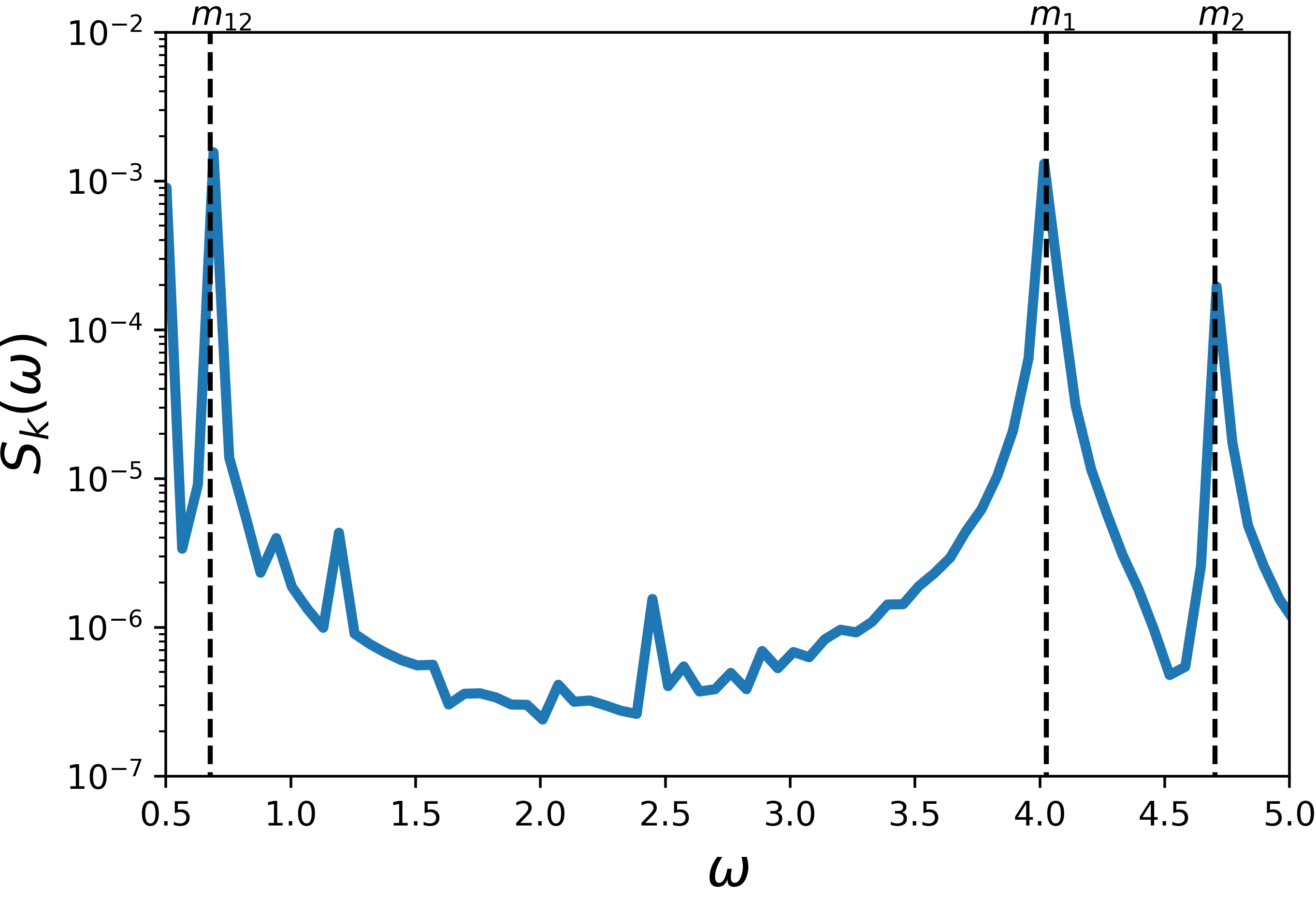}
    \caption{Power spectrum $S_k(\omega)$ of the Krylov complexity for a quench within the ferromagnetic phase to $h_z = 0.2$ and $h_x=0.25$, starting from a fully polarized ferromagnetic state. The high-frequency peaks correspond to the two meson masses $m_1 = 4.025$ and $m_2 = 4.702$ obtained from the semiclassical analysis. The spectrum also captures the relative spacing between meson masses $m_{12}$. We consider a system of size $L = 14$ with Krylov complexity evolved up to time $t = 100$ with a time step of $\Delta t = 0.1$ for computing the power spectrum using DFT.
}
    \label{f5}
\end{figure}

To make this connection explicit, we analyze the dynamics in the frequency domain through the power spectrum of the Krylov complexity:
\begin{equation}
S_k(\omega) = \left|\int_{-\infty}^{\infty} e^{\,i\omega t}\, C_k(t)\, dt \right|^{2}.
\end{equation}
As shown in Fig.~\ref{f5}, $S_k(\omega)$ computed through a discrete Fourier transform (DFT) exhibits distinct peaks at frequencies that match the meson masses predicted by semiclassical analysis. This correspondence demonstrates that the spectral decomposition of Krylov complexity not only captures the confined nature of the excitations but also quantitatively resolves the hierarchy of mesonic masses. In this sense, Krylov complexity acts not merely as a measure of quantum-state spreading under time evolution, but also as a sensitive probe of confinement dynamics, encoding detailed information about the emergent bound-state spectrum in its spectral structure.

Finally, we note that Krylov-based spectroscopy offers several advantages over traditional operator-based approaches. Consider the expectation value of a general operator $\hat{\mathcal{O}}$, $\langle \hat{\mathcal{O}}(t) \rangle = \sum_{a,b} c_a c_b^* e^{-i(E_a-E_b)t} \langle E_b | \hat{\mathcal{O}} | E_a \rangle$, where $|\Psi_0\rangle = \sum_a c_a |E_a\rangle$ is the initial state expanded in the Hamiltonian eigenbasis. The corresponding power spectrum is given by $|\hat{\mathcal{O}}(\omega)|^2 \propto \sum_{a,b} |c_a|^2 |c_b|^2 \delta(\omega - \Delta_{ab}) |\langle E_b|\hat{\mathcal{O}}|E_a\rangle|^2$, showing that any vanishing matrix elements $\langle E_b|\hat{\mathcal{O}}|E_a\rangle$ will cause certain energy differences $\Delta_{ab}=E_a-E_b$ to be missed. In contrast, Krylov state complexity $C_k(t)$ depends only on the initial state and the Hamiltonian, and its dynamics scans all frequency components accessible from the initial state. As a result, no dynamically connected energy levels are overlooked, making Krylov complexity a state-specific, operator-independent probe that faithfully captures the complete spectrum of accessible excitations and provides a robust tool for analyzing the system’s underlying energy structure.

\textbf{\emph{Conclusions.---}}We have demonstrated that Krylov complexity serves as a sensitive probe of confinement dynamics in low-dimensional quantum systems. Focusing on the transverse-field Ising model, we explored its behavior under various quench protocols. For quenches within the ferromagnetic phase, the complexity is strongly suppressed even for small longitudinal fields, consistent with the absence of correlation spreading under confinement. In contrast, without the longitudinal field, the complexity exhibits large-amplitude oscillations associated with freely propagating domain-wall excitations. For quenches within the paramagnetic phase ($h_x > 1$), the Krylov complexity increases with the longitudinal field, reflecting the non-confining nature of this regime. Quenches across the critical point from the paramagnetic to the ferromagnetic phase show a more intricate behavior where the complexity initially grows but eventually decreases with increasing longitudinal field, suggesting the onset of weak confinement. Notably, the overall magnitude of the complexity in this case is several orders of magnitude larger than in the other scenarios, likely due to the quench across the critical point exciting a broad continuum of modes and generating strong delocalization in Krylov space. Finally, we showed that the oscillation frequencies of the complexity in the confining regime match the meson masses, with peaks in its power spectrum aligning precisely with the semi-classically predicted values.

Our results thus establish Krylov complexity as a powerful diagnostic of confinement in low-dimensional quantum systems, bridging concepts from quantum information theory with emergent phenomena traditionally studied in high-energy and condensed-matter physics. Several open questions naturally follow. For instance, how does Krylov complexity behave in lattice gauge theories where confinement emerges from gauge constraints~\cite{Yaffe1980,Kebri2024,CHANDRASEKHARAN1999739,Jesse2025,halimeh2025,Jesse2025,osborne2024,Halimeh2022tuning,Cheng2022,Zhang2023,Mildenberger2025}? Can similar diagnostics capture other nonperturbative phenomena such as false-vacuum decay \cite{Lagnese2021,zhu2024} and string breaking \cite{mwy1-v9hk,Surace2020}. 
Another promising direction is to investigate how confinement affects operator growth in Krylov space, employing finite-temperature algorithms~\cite{TanCheng2025} to probe phenomena close to the ground state.

\bigskip
\footnotesize
\begin{acknowledgments}
The authors acknowledge funding by the Max Planck Society, the Deutsche Forschungsgemeinschaft (DFG, German Research Foundation) under Germany’s Excellence Strategy – EXC-2111 – 390814868, and the European Research Council (ERC) under the European Union’s Horizon Europe research and innovation program (Grant Agreement No.~101165667)—ERC Starting Grant QuSiGauge. Views and opinions expressed are, however, those of the author(s) only and do not necessarily reflect those of the European Union or the European Research Council Executive Agency. Neither the European Union nor the granting authority can be held responsible for them. This work is part of the Quantum Computing for High-Energy Physics (QC4HEP) working group.
\end{acknowledgments}

\clearpage
\widetext
\begin{center}
\textbf{\large Supplemental Material for `Krylov Complexity Meets Confinement'}
\end{center}
\makeatletter
\renewcommand{\c@secnumdepth}{0}
\makeatother
\setcounter{equation}{0}
\setcounter{figure}{0}
\setcounter{table}{0}
\setcounter{page}{1}
\makeatletter
\renewcommand{\theequation}{S\arabic{equation}}
\renewcommand{\thefigure}{S\arabic{figure}}
\renewcommand{\thesection}{S\arabic{section}}

\section{Semi-Classical Approach for Determining Meson Masses}

\begin{figure}[htbp]
    \centering
    \begin{minipage}{0.5\textwidth}
        \centering
        \includegraphics[width=\textwidth]{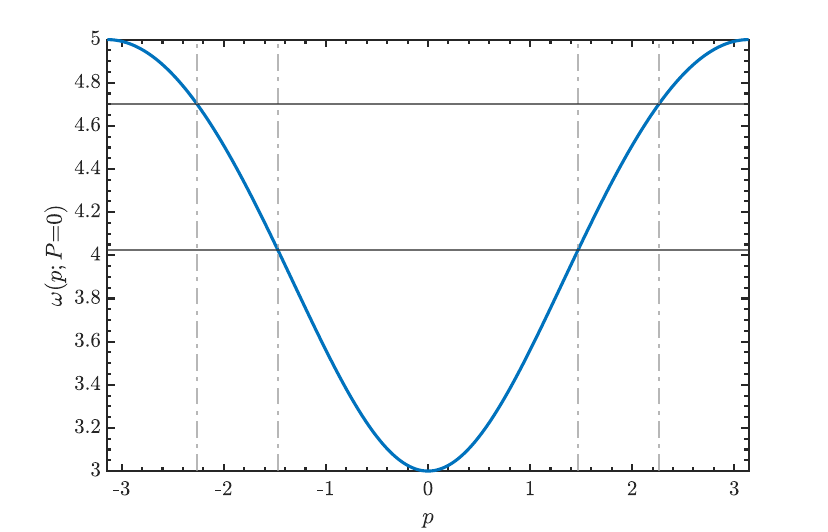}
       (a)
    \end{minipage}
    \hfill
    \hspace{-0.5\textwidth}
    \begin{minipage}{0.5\textwidth}
        \centering
        \includegraphics[width=\textwidth]{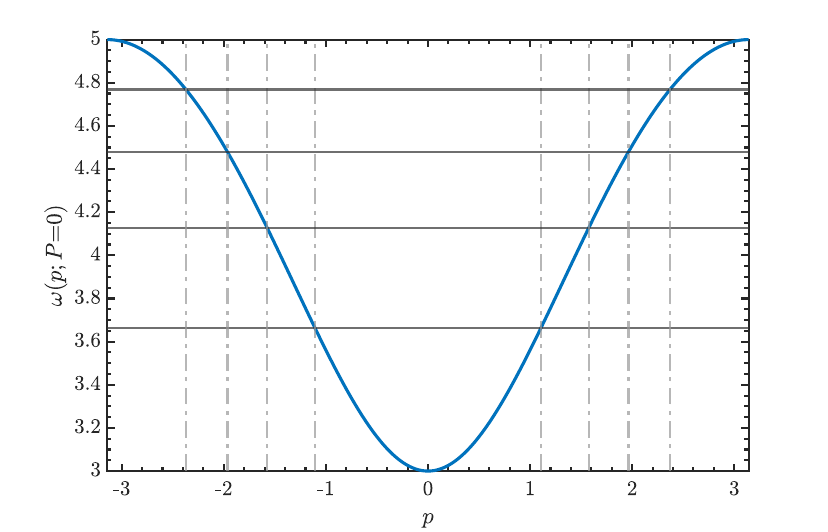}
        (b)
        
    \end{minipage}
    \label{S2}
    \caption{Semiclassical bound-state energy levels in $\omega(p,P)$ obtained from the solutions of Eq.~\ref{BS1}. Dashed vertical lines indicate the turning points $p_{a,b}$. Horizontal lines denote the meson mass values. (a) Bound states for $h_x = 0.25, h_z = 0.2, P = 0$. (b) Bound states for $h_x = 0.25, h_z = 0.1, P = 0$.}
\end{figure}

In the ferromagnetic phase ($h_x < 1$) of the transverse field Ising chain,
the low-energy excitations are domain walls separating regions of opposite magnetization.
When a small longitudinal field $h_z$ is applied, it lifts the degeneracy between the two ferromagnetic ground states.
Domains aligned opposite to the field acquire an energy proportional to their spatial extent,
producing a linear attractive potential between neighboring domain walls.
As a result, isolated domain walls are confined into bound states ``mesons'' that are analogous to those in particle physics. We employ a semiclassical analysis, closely following the computations presented in Refs.~\cite{Kormos2017,Rutkevich2008}.

We describe the dynamics of two neighboring domain walls (treated as fermions) by the semiclassical Hamiltonian
\begin{equation}
H = \varepsilon(p_1) + \varepsilon(p_2) + \chi |x_2 - x_1|,
\end{equation}
where $\varepsilon(p)$ is the single-particle dispersion relation $\varepsilon(p)=2J\sqrt{1 - 2h_x \cos p + h_x^2}$ and the confining strength is given by $\chi = 2J h_z\overline{\sigma}$, where 
$\overline{\sigma} = (1 - h_x^2)^{1/8}$.

Introducing the center-of-mass and relative variables $X = \frac{x_1 + x_2}{2},  x = x_2 - x_1, 
P = p_1 + p_2, p = \frac{p_2 - p_1}{2}$, the Hamiltonian becomes
\begin{equation}
H = \omega(p;P) + \chi |x|, \qquad
\omega(p;P) = \varepsilon\!\left(p + \tfrac{P}{2}\right) + \varepsilon\!\left(p - \tfrac{P}{2}\right).
\end{equation} The Hamiltonian equations of motion read
\begin{equation}
\dot{X} = \frac{\partial \omega}{\partial P}, \qquad
\dot{x} = \frac{\partial \omega}{\partial p}, \qquad
\dot{p} = -\chi\,\mathrm{sgn}(x), \qquad P(t) = \text{const}.
\end{equation}.

The quantized bound-state energies $E_n(P)$ follow from the Bohr--Sommerfeld condition. The potential $\omega(p_a;P)$ has a single minima at $p=0$ when $P < 2\arccos h_x$ and in this case the semi-classical energy levels are given by,
\begin{equation}\label{BS1}
2E_n(P)p_a - \int_{-p_a}^{p_a}\!dp\,\omega(p;P)
 = 2\pi\chi\!\left(n - \tfrac{1}{4}\right),
\end{equation} where $p_a$ are the turning points that satisfy $\omega(p_a;P) = E_n(P)$. When $P >2\arccos h_x$, the potential has two minima with the energy levels given by,

\begin{figure}[H]
\centering
    \includegraphics[width=10cm]{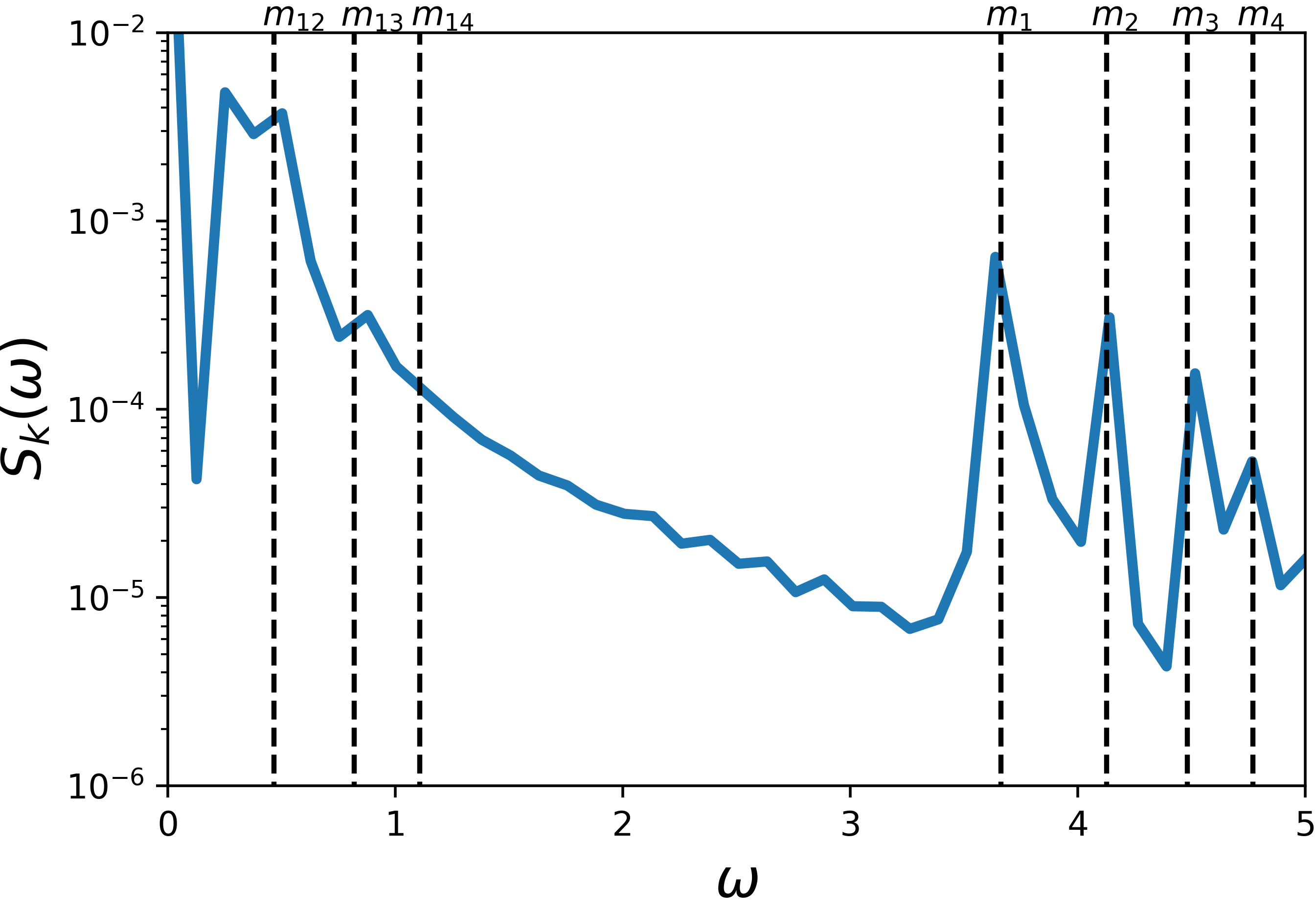}
    \caption{Power spectrum $S_k(\omega)$ of the Krylov complexity for a quench within the ferromagnetic phase to $h_z = 0.1$ and $h_x=0.25$, starting from a fully polarized ferromagnetic state. The high-frequency peaks correspond to the four meson masses $m_1 = 3.662$, $m_2 = 4.127$, $m_3 = 4.48$ and $m_4=4.769$ obtained from the semiclassical analysis. The spectrum also captures approximately the relative spacing between meson masses such as $m_{12}$ and $m_{13}$. We consider a system of size $L = 14$ with Krylov complexity evolved up to time $t = 50$ with a time step of $\Delta t = 0.1$ for computing the power spectrum using DFT.
}
    \label{S1}
\end{figure}

\begin{equation}\label{BS2}
E_n(P)(p_a - p_b) - \int_{-p_b}^{p_a}\!dp\,\omega(p;P)
 = \pi\chi\!\left(n - \tfrac{1}{2}\right),
\end{equation}
with turning points satisfying $\omega(p_{a,b};P)=E_n(P)$.
For the quenches within the ferromagnetic phase shown in Fig.~1 of the main text,
corresponding to parameters $h_x = 0.2$ and $h_z = 0.25$,
we numerically solve Eq.~\eqref{BS1} and identify two mesonic bound states
with masses $m_1 = 4.025$ and $m_2 = 4.702$, as illustrated in Fig.~\ref{S2}(a). Upon reducing the longitudinal field to $h_z = 0.1$, the solution admits four confined states with masses $m_1 = 3.662$, $m_2 = 4.127$, $m_3 = 4.480$, and $m_4 = 4.769$, as shown in Fig.~\ref{S2}(b). The peaks observed in the power spectrum of the Krylov complexity $S_k(\omega)$, displayed in Fig.~\ref{S1}, successfully reproduce these meson masses with good accuracy.

\section{Lanczos algorithm}

To compute the Lanczos coefficients associated with a given quantum state, we employ the Full Orthogonalization Lanczos (FOL) procedure (see \cite{Caputa:2025ucl,Rabinovici:2020ryf} for a detailed review). Starting from an initial normalized state $|K_0\rangle$, the algorithm iteratively constructs an orthonormal Krylov basis $\{|K_n\rangle\}$ through successive applications of the Hamiltonian, $\hat{H}$ employing Gram--Schmidt orthogonalization procedure as shown below. The FOL is one version of the Arnoldi iteration and when $\hat{H}$ is Hermitian, it automatically reduces to the standard Lanczos algorithm. In the FOL variant, each new vector is orthogonalized against \emph{all} previously generated basis vectors to avoid numerical loss of orthogonality, ensuring high precision in the computed coefficients even for long-time dynamics. We terminate our algorithm when we hit the maximum dimension of the Krylov space by choosing a cutoff when the Lanczos coefficients $\beta$ satisfy $\beta_n<\epsilon$ with $\epsilon=10^{-12}$. The resulting sequence $\{\alpha_n, \beta_n\}$ characterizes the propagation of the initial state in Krylov space and forms the foundation for evaluating the Krylov complexity. We give the detailed steps involved in the algorithm below.

\begin{algorithm}[H]
\caption{Full Orthogonalization Algorithm}
\begin{algorithmic}[1]
\State \textbf{Initialization:}
\Statex \hspace{1em} $|K_0\rangle = \frac{1}{\sqrt{\langle\Psi_0|\Psi_0\rangle}}|\Psi_0\rangle$
\State \textbf{Initial Parameters:}
\Statex \hspace{1em} $\alpha_0 = \langle K_0|H|K_0\rangle$, \quad $\beta_0 = 0$
\For{$n \ge 1$}
    \State $|A_n\rangle = H|K_{n-1}\rangle$
    \State \textbf{Gram--Schmidt orthogonalization:}
    \Statex \hspace{1em} $|A_n'\rangle = |A_n\rangle - \sum_{i=0}^{n-1}|K_i\rangle\langle K_i|A_n\rangle$
    \State Repeat Gram--Schmidt for numerical stability
    \State $\beta_n = \sqrt{\langle A_n'|A_n'\rangle}$
    \If{$\beta_n < \epsilon$}
        \State \textbf{break}
    \Else
        \State $|K_n\rangle = \frac{1}{\beta_n}|A_n'\rangle$
        \State $\alpha_n = \langle K_n|H|K_n\rangle$
    \EndIf
\EndFor
\end{algorithmic}
\end{algorithm}

\end{document}